\documentclass[twocolumn,pre,superscriptaddress]{revtex4} 
\usepackage{graphicx,float} 
\usepackage{latexsym} 
\usepackage{epstopdf} 
\usepackage{amssymb,amsmath} 
\usepackage{color}
\usepackage{multirow} 
\usepackage{dsfont} 
\usepackage{hyperref} 
\usepackage{graphicx} 
\makeatletter \let\Hy@linktoc\Hy@linktoc@none \makeatother
\hypersetup{colorlinks=true,linkcolor=blue,citecolor=red} 
\usepackage{verbatim} 
\usepackage{bm} 
\begin{document} 
\title{Fractional Stochastic Loewner Evolution and Scaling Curves}
\author{M. Ghasemi Nezhadhaghighi} 
\email{mgnhaqiqi@gmail.com} 
\affiliation{Department of Physics, Faculty of Science, Shiraz University,
Shiraz 71454, Iran} \date{\today}

\begin{abstract} 

The Stochastic Loewner equation, introduced by Schramm, gives us a powerful way to study and classify critical random curves and interfaces in two-dimensional statistical mechanics. New kind of stochastic Loewner equation, called fractional stochastic Loewner evolution (FSLE), has been proposed for the first time. Using the fractional time series as the driving function of the Loewner equation and local fractional integrodifferential operators, we introduce a large class of fractal curves. We argue that the FSLE curves, besides the fractal dimension calculations, have crucial differences which caused by the Hurst index of the driving function. This extension opens a new way to classify different types of scaling curves based on the Hurst index of the corresponding driving function. Such formalism appear to be suitable to deal with the study of a wide range of two-dimensional curves appearing in statistical mechanics or natural phenomena.

\end{abstract} 

\maketitle

\section{Introduction}
Studying the critical interfaces in two dimensions is in the interest of both physicists and mathematicians because of at least two reasons: firstly
because of application in the physical systems such as domain walls in statistical models \cite{bernard0}, iso-hight lines in rough surfaces
\cite{rajab}, zero-vorticity lines of Navier-Stokes turbulence \cite{bernard}, the continuum limit watersheds dividing drainage basins \cite{hermaan2012}, 
iso-height contour lines of suspended graphene sheets \cite{hermaan2016} and the two-dimensional projection of cumulus clouds \cite{hermaan2021}. 
Secondly these interfaces when they posses conformal symmetry can be
studied by the exact methods of conformal field theory (CFT)\cite{BPZ} and rigorous methods of Schramm-Loewner evolution (SLE)\cite{schramm}. Both of
the above methods are based on the conformal symmetry of the interfaces which is the property of the most of the systems at the critical point in two
dimensions. 

However, it seems that systems with just scaling symmetry and not conformal symmetry is also ubiquitous, some examples are rough surfaces
\cite{rajab2}, two-dimensional theory of elasticity \cite{Riva cardy}, fractures in highly disordered materials \cite{hermaan20121}, percolation with 
long-range correlated disorder \cite{hermaan2013}, perimeters of multilayered and directed percolation clusters \cite{hermaan20162}, and off-critical statistical 
systems \cite{shibasa2020}.

\begin{figure} 
\begin{center}
\includegraphics[angle=0,scale=0.3]{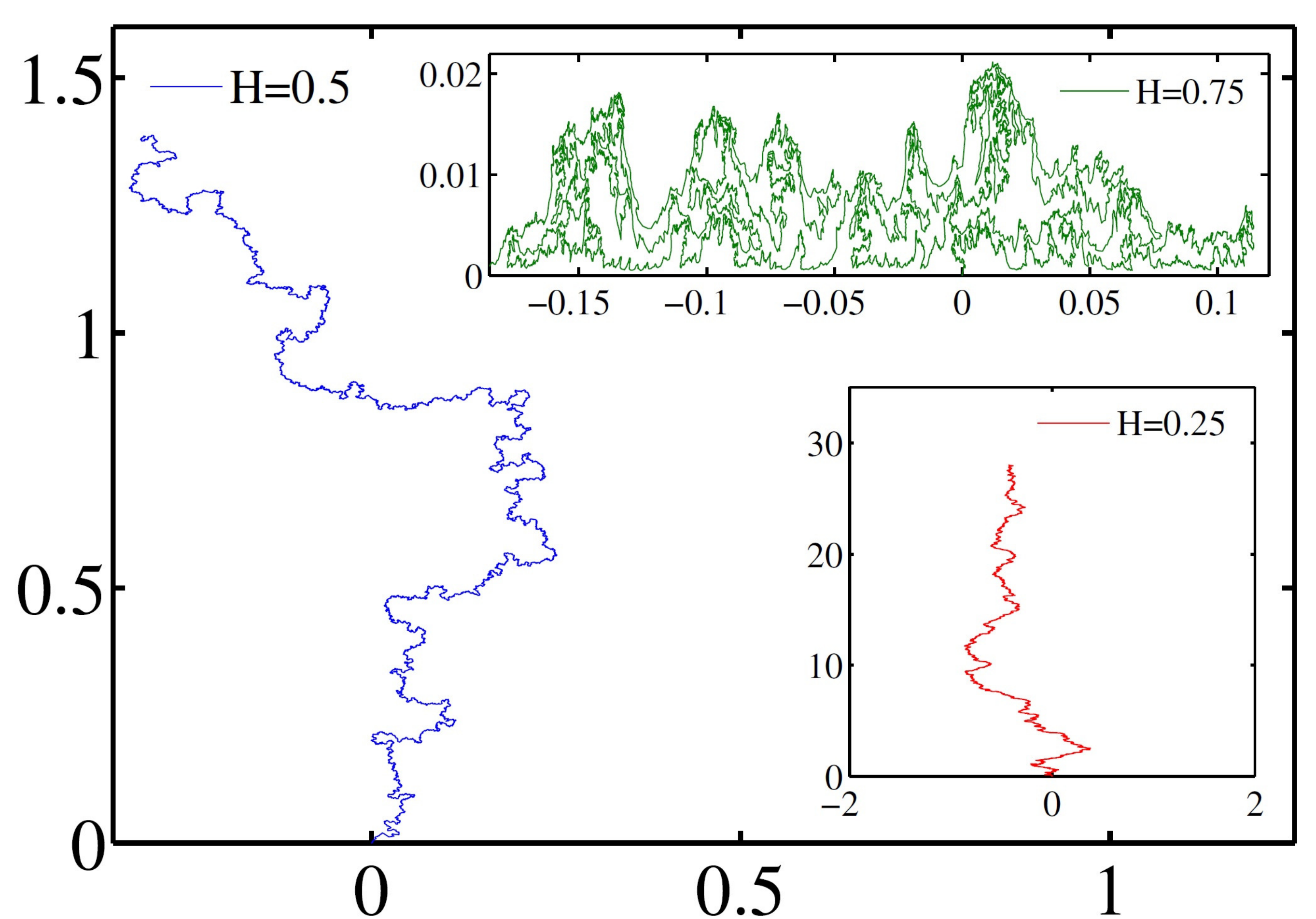}\\ 
\caption{FSLE trace for FBM with $\kappa=2$, $71000$ steps for $H=0.25$ and $H=0.5$
and $30000$ steps for $H=0.75$.} \label{Fig1} 
\end{center} 
\end{figure}

The non-locality or non-unitarity of the interaction are necessary to have a
system with such behavior. One way to study these systems is by using Loewner equation similar as the ordinary SLE but with drift which is now just
scale invariant and not conformal invariant. Schramm-Loewner evolution is based on describing curves in the upper half plane $\mathcal{H}$ by using a map 
$g_{t}(z)$ which is
mapping upper half plan minus hull to the upper half plane. By hull we mean the curve and the region of the upper half plane which is separated from
infinity by the curve. This map satisfies the Loewner equation 
\begin{eqnarray}\label{SLE}
\partial_{t}g_{t}(z)=\frac{2}{g_{t}(z)-U_{t}}
\end{eqnarray}
with initial condition $g_0(z) = z$. Note that $U_{t}$ as a drift of the process can
be any continuous real function of time, which will be called driving function. For ordinary SLE which is describing conformal Markov curves $U_t$ is chosen to be the rescaled Brownian motion $\sqrt{\kappa}B_t$ (Brownian motion is a Gaussian Markov stochastic process with mean 
zero and $E[B_tB_s = \min(t,s)$), then the hulls are fractal curves, whose properties depend on $\kappa$
 (for instance, their fractal dimension is $d_f = \min(1+\frac{\kappa}{8},2) $) \cite{beffara}.  

In this paper we would like to use fractional
Brownian motion and fractional Weierstrass function as drifts of the stochastic Loewner equation. Such a generalization is not new, SLE with L\'evy drift 
was already discussed in both physics and
mathematics literature, see \cite{Gruzberg2,guan}. Since the introduced curves in these papers are not continues and not scale invariant they are not
useful to describe continues scale invariant objects. 

Although several researches have been suggested that the corresponding driving function to some fractal curves,  display very distinctive anomalous diffusive 
 \cite{hermaan20162} or chaotic \cite{shibasa2020} behaviors.  

In this paper we would like to study SLE with fractional Brownian motion (FBM) and also fractional  Weierstrass function as a drift. Fractional Brownian motion is
the continues generalization of Brownian motion and has scale invariancy. The Weierstrass function is a real-valued function that
 is continuous everywhere but differentiable nowhere. It is an example of a fractal time series. To make SLE appropriate for these processes we will introduce a new
parametrization for the SLE curve to produce a scale invariant hull. To study the special properties of the curve such as the probability of hitting
the real line and fractal properties of the scaling curves, we use qualitative arguments in the spirit of our numerical experiment. The paper organized as follows: 
first we will briefly review
FBM and fractional Weierstrass function and then we will introduce fractional SLE. Then we will study some global properties of the fractional SLE curves such as the 
hitting the boundary and fractal properties. We will confirm our qualitative arguments by extensive numerical experiments. Finally in Section we give the main conclusions of our work.

\begin{figure} 
\begin{center}
 \includegraphics[angle=0,scale=0.3]{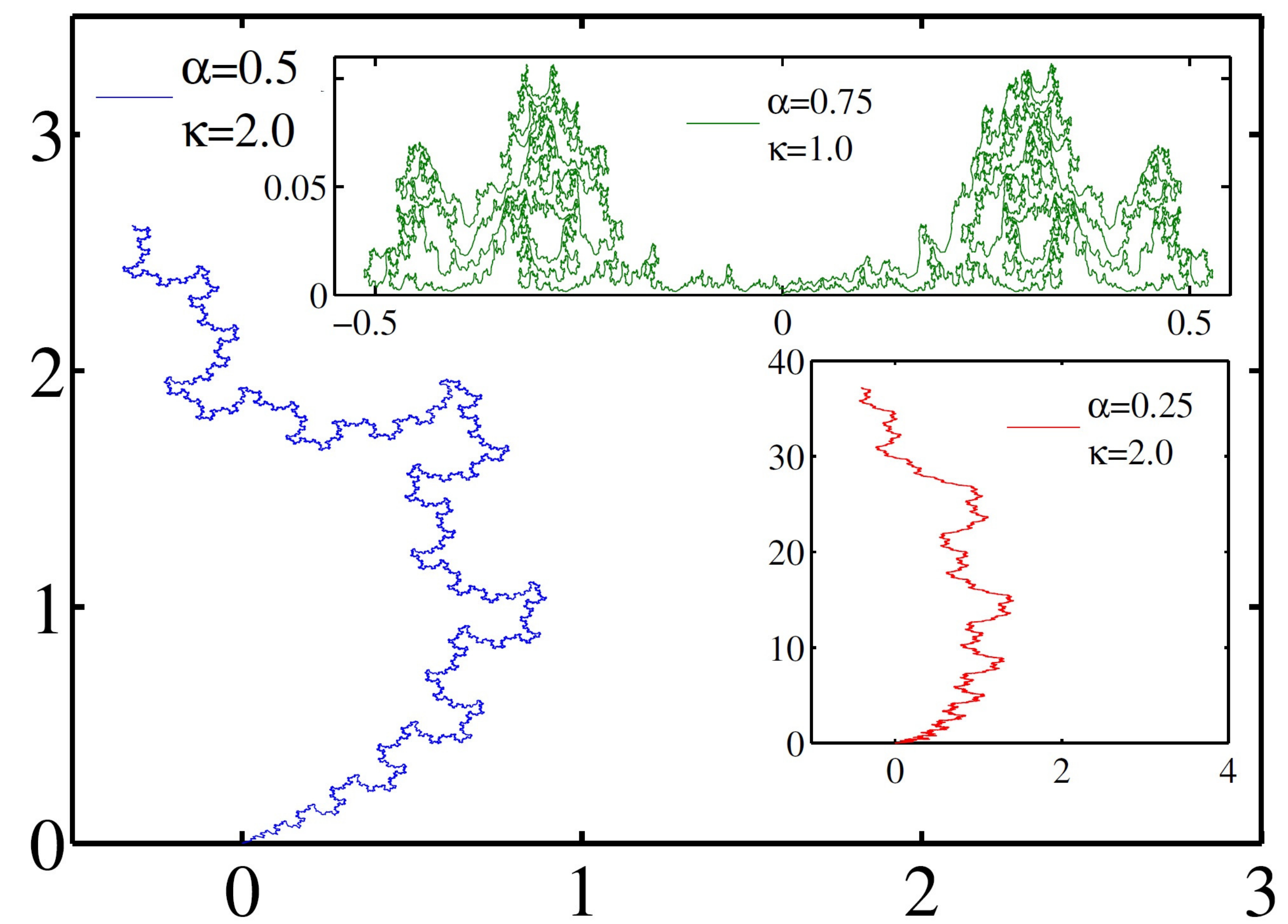}\\ 
\caption{FSLE trace with
fractional Weierstrass-Mandelbrot (FWM) function as a drift with 60000 steps (in all plots $\alpha = H$).} \label{Fig2} 
\end{center} 
\end{figure}

\section{Fractional time series} 
This article develops a new version of stochastic Loewner evolution, obtained by taking as driving function a fractional time series. Here we use two different ways to generate a time series with the desired fractional properties. The first is fractional Brownian motion and the latter is fractional Weierstrass-Mandelbrot function. In this section, we discuss the properties of these variations of fractional time series.  

\textit{$I$. Fractional Brownian motion:}\\

There are different representations for fractional Brownian motion, in this paper we will use the definition by Mandelbrot and Van Ness \cite{MV}, 
\begin{eqnarray}\label{FBM}
B_{t}^{H}=\frac{1}{\Gamma(H+\frac{1}{2})}\{\int_{-\infty}^{0}((t-s)^{H-\frac{1}{2}}-(-s)^{H-\frac{1}{2}})dB_{s} \nonumber \\
+\int_{0}^{t}(t-s)^{H-\frac{1}{2}}dB_{s}\} ,
\end{eqnarray}
where $B_{s}$ is the normal Brownian motion and $0<H\leq1$. The $H=1$ has a very simple form and it is called ballistic motion $B_{t}^{H}=\xi t$,
where $\xi$ is the random Gaussian noise. Fractional Brownian motion is not a Markov process, however, it does have stationary increments. 
The autocorrelation is
\begin{eqnarray}
\langle B^{H}_{t+T}B^{H}_{t}\rangle=\frac{1}{2}((t+T)^{2H}+t^{2H}-T^{2H}).
\end{eqnarray}
For $T=0$, this implies anomalous diffusion with mean-square-displacement $\langle B_H^2(t)\rangle = t^{2H}$, where the anomalous diffusion exponent $\alpha = 2H$ 
can take values between $0$ and $2$. In the sub-diffusive case, $0 < \alpha < 1$, the increments are anti-correlated
(anti-persistent) while the motion is persistent (positive correlations between the steps) in the super-diffusive case $1 < \alpha < 2$ \cite{Wada}.

The FBM admits a version whose sample paths are almost surely H\"{o}lder continuous of
order strictly less than $H$. We should mention that up to the level of numerical calculations most of the forthcoming arguments are also valid for the
generic drifts with scaling property, such as, non-stationary FBM, Levy processes and continues time random walk \cite{Metzler}.
\\
\\
\textit{$II$. Fractional Weierstrass-Mandelbrot function:}\\

The fractional Weierstrass-Mandelbrot (FWM) function, is a random continuous
non differentiable mono fractal function. The discrete scale invariance (DSI) properties of FWM function which was studied by Berry and Lewis in \cite{weierstrass},
 is the key difference between FWM and FBM time series. A wide range of physical phenomena such
as sediment and turbulence \cite{Sornette1,Sornette2}, fractal properties of the landscapes and other enviromental data \cite{Burrough}, contact analysis of elastic-plastic fractal surfaces \cite{Yan} and propagation
and localization of waves in fractal media \cite{Garcia}, could be modeled by the Weierstrass-Mandelbrot (WM) function.

The FWM function which is not Markov but has stationary increments can be defined as follows
\begin{eqnarray}\label{FWM}
W(h,t) = \sum_{n=-\infty}^{\infty} \lambda ^{-nH}(h(0)-h(\lambda^n t))e^{i\phi_n}
\end{eqnarray}
where $h(t)$ can be any periodic function which is differentiable at zero and $\phi_n$ is a random phase chosen from the interval $[0,2\pi]$ uniformly, 
to make the series convergent we need to consider $0 < H < 1$ ($H$ is the Hurst index of the process and the fractal dimension of the process is $2-H$). 
It was proved in \cite{weierstrass,Szulga} that the
above process with $h(x) = e^{ix}$ converges to the fractional Brownian motion in the limit $\lambda \to 1$. In our numerical
calculation we will take $h(x) = e^{ix}$, however, our results are extendable to the most generic cases \cite{nezhadgaghighi}.

\section{Fractional Stochastic Loewner Evolution}

In recent years, growing attention has been focused on the processes that
take place in random disordered environments \cite{Bressloff}. As a result of this highly irregular processes, the anomalous properties of transport processes, occurs. Fractional calculus, (for example the appearance of the fractional time derivative by a simple change $\frac{\partial}{\partial t}\to \frac{\partial^\alpha}{\partial t^\alpha}$), is a powerful mathematical tool for the investigation of these kinds of physical phenomena such as motion in disordered media \cite{Barkai2008}. The history of fractional calculus began in 1695 by Leibniz and after
that several mathematicians have defined fractional order derivatives. The well known definition is the Riemann–Liouville, which can be derived
considering the so-called differintegral operators \cite{hilfer}. The Riemann–Liouville definition is as follows:
\begin{eqnarray}\label{RLO}
\frac{d^{\alpha}f(t)}{[d(t-t_0)]^{\alpha}}=\frac{1}{\Gamma(1-\alpha)}\frac{d}{dt}\int_{t_0}^{t}\frac{f(t^\prime)}{(t-t^\prime)^{\alpha}}dt^\prime. 
\end{eqnarray}
It is obvious, the operator ${d^{\alpha}}/{[d(t-t_0)]^{\alpha}}$ is non-local and further ${d^{\alpha}f(t)}/{[d(t-t_0)]^{\alpha}}\neq 0$ for $f(t) =$const. In order to correct this feature a new notion of local fractional derivative
(LFD) was introduced by \cite{Kolwankar}. The definition of the left-sided local fractional derivative of order $\alpha$ was defined by
\begin{equation}\label{LFD}
\mathbf{D}^{\alpha}f(t)=\lim_{{t\rightarrow
t^\prime}}\frac{d^{\alpha}[f(t)-f(t^\prime)]}{[d(t-t^\prime)]^{\alpha}}
\end{equation}
where the operatore in the RHS is the Riemann-Liouville derivative (\ref{RLO}). An easy calculation shows that
\begin{eqnarray}\label{DLFD}
\mathbf{D}^{\alpha}f(x)=\Gamma(1-\alpha)\lim_{_{\Delta\rightarrow0}}\frac{f(x-\Delta)-f(x)}{\Delta^{\alpha}}.
\end{eqnarray}

If one choose fractional Brownian motion or fractional Weierstrass-Mandelbrot function as a drift of normal stochastic Loewner evolution (\ref{SLE}), then the result is just a simple curve with fractal dimension equal to one for $H>\frac{1}{2}$ and it is
space filling for $H<\frac{1}{2}$ (see ref. \cite{nattaghfsle}). This is consistent with the Marshall-Rohde theorem which states that if the drift is better than H\"{o}lder-1/2,
then it defines a smooth (in some appropriate sense) slit \cite{MR}. To make SLE scale invariant under the fractional drift we shall choose new
parametrization for the equation by changing the capacity of the conformal maps, i.e. $cap(K_{t}):=c$, as $dc=2(dt)^{2H}$. One can write the SLE
equation in the new parametrization as follows
\begin{eqnarray}\label{New SLE} 
\mathbf{D}^{2H}g_{t}(z)=\frac{2}{g_{t}(z)-\sqrt{\kappa}B_{t}^{H}},
\end{eqnarray} 
where we used local fractional derivative (\ref{LFD}). If one is further interested in computing the discretized version of above equation (see \ref{DLFD}), one can simply  use
\begin{eqnarray}\label{New SLE} 
dg_{t}(z)=\frac{2(dt)^{2H}}{g_{t}(z)-\sqrt{\kappa}B_{t}^{H}}.
\end{eqnarray} 
We call the above equations, fractional
Stochastic Loewner evolution (FSLE), the velocity of the growing curve is different from the ordinary SLE. 

The equation (\ref{New SLE}) has the scaling property in
the following sense: the conformal map $\tilde{g}_{t}(z)=\frac{1}{\alpha^{4H^{2}}}g_{\alpha^{4H}t}(\alpha^{4H^{2}}z)$ with
$\tilde{B}^{H}(t):=\frac{1}{\alpha^{4H^{2}}}B^{H}(\alpha^{4H}t)$ satisfies the same Loewner equation as (\ref{New SLE}). Using the above scaling
property one could claim that $\tilde{g}_{t}(z)$ is the conformal map that takes the complement of the hull
$\tilde{K}_{t}:=K_{\alpha^{4H}t}/\alpha^{4H^{2}}$ on to the half plane in the way that $cap(\tilde{K}_{t})=2(dt)^{2H}$. In other words by using the
new parametrization we generated scaling conformal maps with scaling hulls. If we take the FBM or FWM with stationary increments as a drift then it is
possible to generate also stationary FSLE, see Figs. (\ref{Fig1}) and (\ref{Fig2}). The important point to get simple curves for a fractal drift is matching the power of $dt$
with the fractal dimension of the drift. For example for a fractal drift, deterministic or stochastic, with fractal dimension $d_{f}$ we need to
consider $(dt)^{4-2d_{f}}$ as a parametrization of FSLE, any other choice will lead to a space filling hull or trivial curve with fractal dimension one.
\section{Results}
In this section, we investigate the statistical properties of FSLE curves. More precisely, we will study the problem of hitting time the real line which shows a special properties and then we will study the fractal properties of the scaling curves generated by fractional stochastic Loewner evolution. 
\subsection{The problem of hitting the real line} 

One of the simplest properties of the above stochastic curves is the probability of hitting the real
line. As it is well known from the SLE literature the problem of hitting the real line is equivalent to calculating the probability that one arbitrary
point $x$ on the real line is part of the hull after finite time or not. This is equivalent to the probability that the process
$f_{t}(x)=g_{t}(x)-\sqrt{\kappa}B_{t}$ hits the origin \cite{schrammrhode}. A crude way to investigate this problem is to look at the drift and stochastic terms of the equation
\begin{equation}
df_{t}(x)=\frac{2(dt)^{2H}}{f_{t}(x)}-\sqrt{\kappa}dB_{t}^{H},
\end{equation}
if the drift term in RHS dominates then the curve will not touch the
real line otherwise it will touch the real line, let us review first the case $H=\frac{1}{2}$. For $\kappa\leq 4$ the drift term will dominate and for
$\kappa>4$ the Brownian motion term so any change in the power of $dt$ in the SLE, like
$df_{t}(x)=\frac{2(dt)^{1+\alpha}}{f_{t}(x)}-\sqrt{\kappa}dB_{t}^{\frac{1}{2}}$ will affect crucially the global properties of the curve. For positive
$\alpha$ the curve will touch the real line for any value of the $\kappa$ and for negative $\alpha$ it will not touch the real line, hull is smooth
curve. The fractal dimension of the hull for positive $\alpha$ is two and for negative $\alpha$ is one. The similar argument is tractable also for
FSLE. For small $\delta t$ the $H$ order of the fractional derivative of $B_t^H$ is independent of the Hurst exponent, i.e. 
\begin{eqnarray}
\frac{B_{t+\delta
t}-B_{t}}{(\delta t)^{H}}\approx \mathcal{O}((2\ln|\ln \delta t|)^{1/2}).
\end{eqnarray}
Roughly one can write 
\begin{eqnarray}
\frac{df_{t}(x)}{(\delta
t)^{H}}=\frac{2(dt)^{H}}{f_{t}(x)}-\sqrt{\kappa}\eta(t),
\end{eqnarray}
where $\eta(t)$ is independent of $H$. Since $H=\frac{1}{2}$ is the critical value for
$H<\frac{1}{2}$ the first term will dominate and so the curve will not touch the real line and for $H>\frac{1}{2}$ the curve will touch the real line
independent of the value of $\kappa$. 

\begin{figure} 
\begin{center}
\includegraphics[angle=0,scale=0.3]{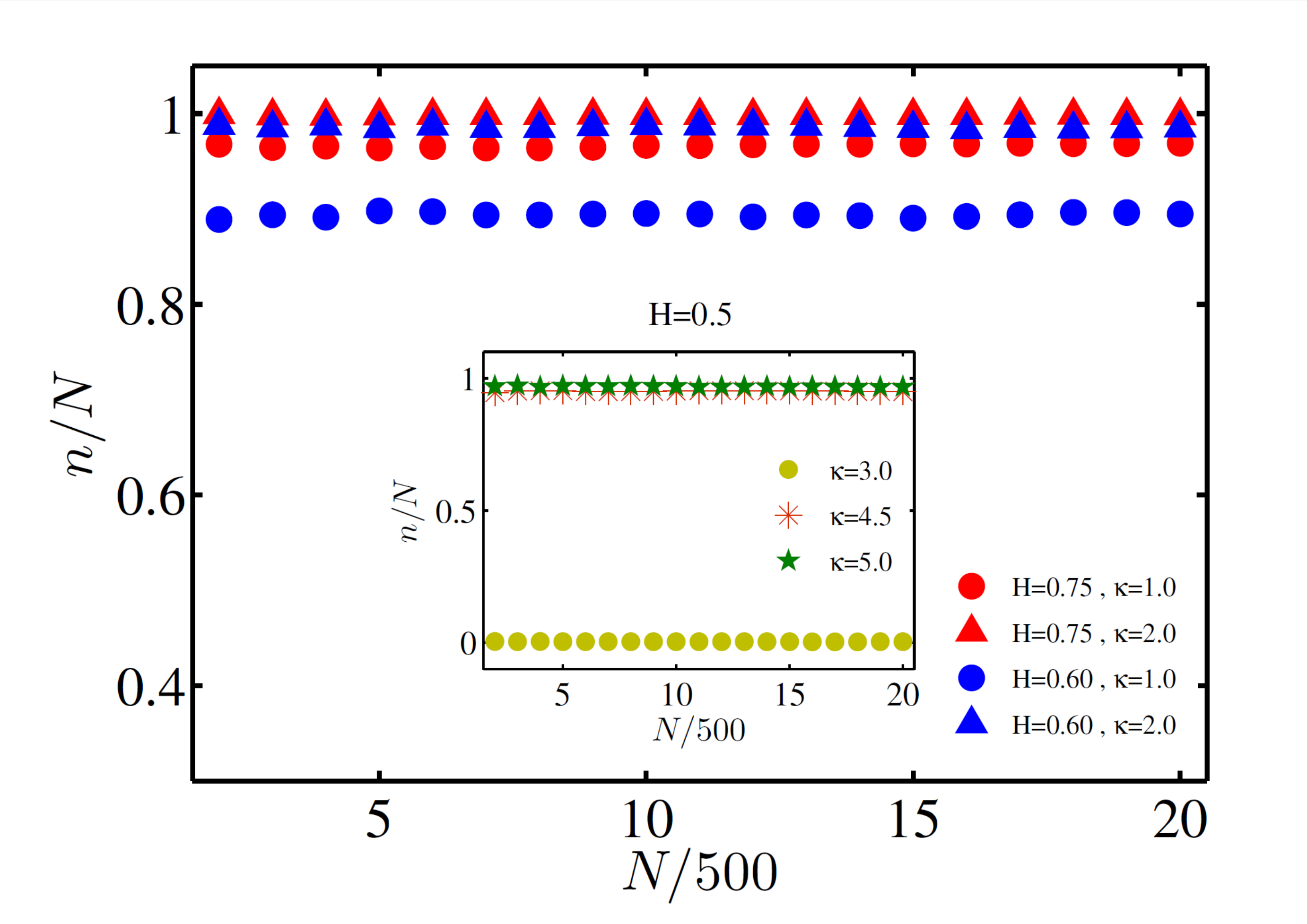}\\
\caption{The probability of touching the real line $n/N$ with respect to the number of realizations $N$ for the curves with 30000 points, in the
inset: the same quantity for the ordinary SLE. }
\label{Fig3} 
\end{center} 
\end{figure}

To get more insight on the rule of the fractional differential one can look at the method of the production of
the hull in the numerical experiment. The trace can be calculated numerically as an iteration process of infinitesimal conformal mappings as follows

\begin{eqnarray}\label{iteration} 
\gamma(j\tau)=f_{1}\circ f_{2}\circ...\circ f_{j}(\xi_{j}), 
\end{eqnarray}
where $\xi_{j}$ is the FBM (FWM) and
\begin{eqnarray}
f_{j}(z)=\sqrt{(z-\xi_{j})^{2}-4\tau^{2H}}+\xi_{j}. 
\end{eqnarray}
The discretized driving force is constant in the interval $[(j-1)\tau,j\tau]$ where
$\tau=\frac{t}{N}$, $N$ is the number of points on the curve, is the mesh of the time. The most simple case is may be the constant force with
$\xi_{j}=0$ then one can get $\gamma_{t}=2\sqrt{t}N^{\frac{1}{2}-H}$. By this equation one can expect that for $H>\frac{1}{2}$ in the numerical
calculation the curve will grow vertically faster than normal SLE. For $H>\frac{1}{2}$ the curve will not reach to infinity in finite time. Of course
the above curves are trivial curves in the limit of large $N$, as explained in \cite{Kolwankar} we just expect non-trivial solution for the equation
(\ref{New SLE}) in the continuum limit when the drift of FSLE has fractal support.

To check numerically the probability of touching the real line we used the above iterative method and countered
the number of curves that are touching the line $y=\delta t$, where $\delta t$ is the average distant between successive discrete points on the curve
as a measure of minimum length. The results  for the probability of touching the real line, for the case when we use FBM as a drift of FSLE, were shown in Fig. \ref{Fig3}.

\subsection{Fractal properties of FSLE curves}

Ordinary SLE (\ref{SLE}) is a Loewner evolution with $U(t) = \sqrt{\kappa}B_t$, where $B_t$ is Brownian motion. The $\kappa$ parameter in the theory,  which is the diffusivity of the
Brownian motion, controls the scaling and fractal properties of the SLE curves. As we mentioned earlier SLE$_\kappa$ is a random fractal curve with the
fractal dimension
\begin{eqnarray}
d_f = 1+\frac{\kappa}{8} \hspace{1cm}(0<\kappa < 8),
\end{eqnarray} 
where it becomes space-filling for $\kappa \geq 8$ with fractal dimension $d_f = 2$ \cite{schramm}. Although SLE gives a powerful way to classify fractal (scale invariant) curves in two dimensions but it was shown in Ref. \cite{nezhadgaghighi} that the Weierstrass-Mandelbrot (WM) function as the drift of the Loewner equation
introduces a large class of fractal curves with discrete scale invariance (DSI).  Based on the extensive numerical simulation in Ref. \cite{nezhadgaghighi}, it was shown that, taking $\sqrt{\kappa} \Re(W(h,t))$ (see Eq. (\ref{FWM}) with Hurst index $H=\frac{1}{2}$) as the drift in Eq.(\ref{SLE}), the solution of WSLE is always a fractal curve with fractal dimension
\begin{eqnarray}
d_f = 1+\frac{\kappa_{eff}}{8},
\end{eqnarray} 
where $\kappa_{eff} = \kappa\frac{1}{2\ln \lambda} \int_0^\infty \vert  \frac{h(0)-h(x)}{x}\vert ^2 dx$.

\begin{figure} 
\begin{center}
\includegraphics[angle=0,scale=0.35]{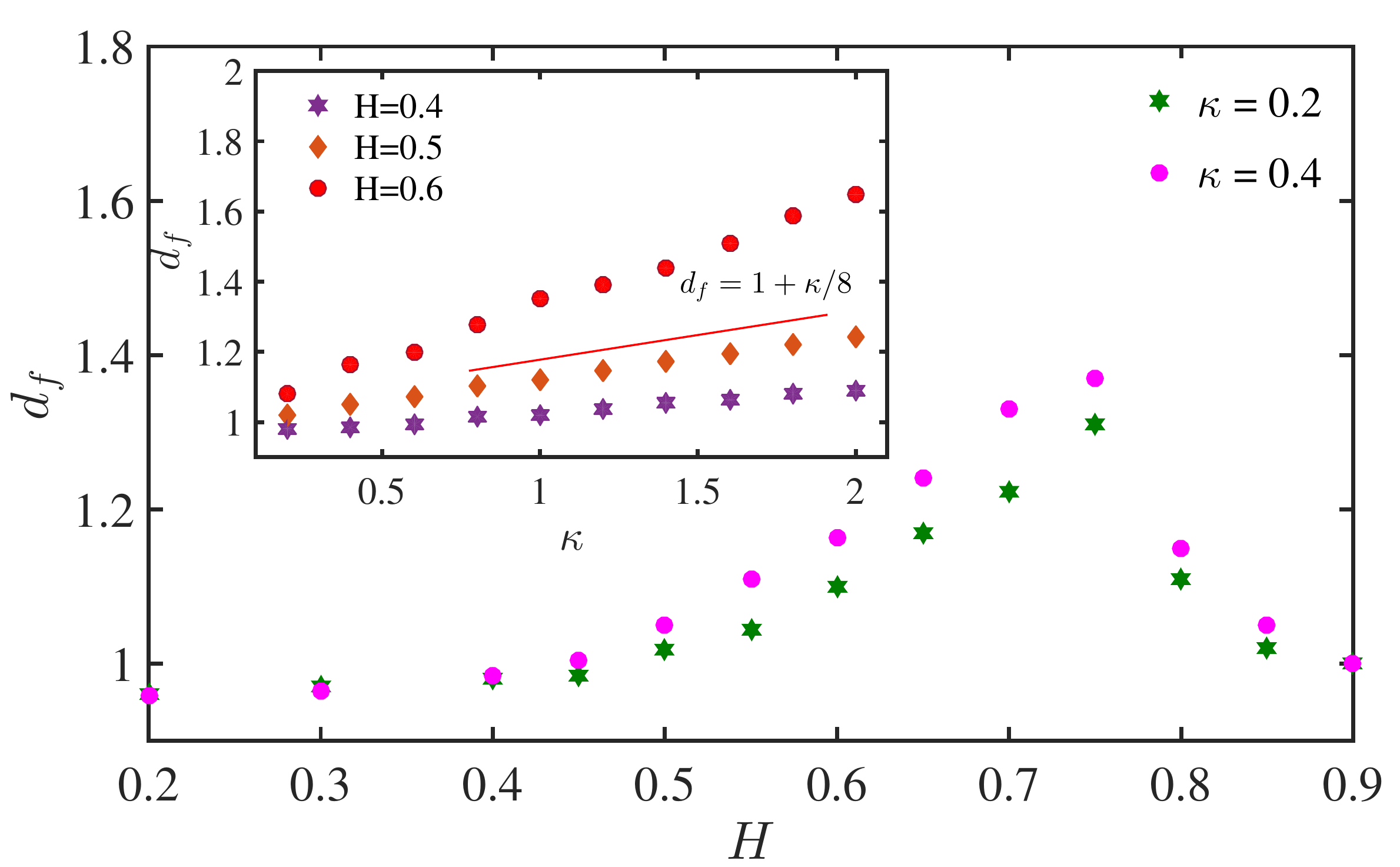}\\
\caption{Fractal dimension $d_f$ of FSLE curves with 30000 points as function of $H$. The same quantity as function of $\kappa$ is depicted in the inset. The most interesting point about this result is the linear dependence of $d_f$ versus $\kappa$ for different values of the Hurst index $H$. }
\label{Fig4} 
\end{center} 
\end{figure}

In this work, we study the scaling and fractal properties of fractional SLE curves in two-dimensions. Let us analyze the case in which the drift in Eq. (\ref{New SLE}) is FBM or FWM (see Eq. (\ref{FBM}) or (\ref{FWM}), respectively). Note that, in order to use the FWM function as a drift of FSLE, we rescale it by the coefficient $\left[\frac{1}{2\ln \lambda} \int_0^\infty \vert  \frac{h(0)-h(x)}{x}\vert ^2 dx\right]^{-1/2}$. This allows us to generate the model-independent scaling curves (the scaling exponents will be independent of $\lambda$ and $h(x)$).    

The total length of the fractal curve, $S$, measured in units of the lattice spacing, $a$, obeys a scaling relation as a function of system size, $L$, as 
\begin{eqnarray}
S\sim \left( \frac{L}{a} \right) ^{d_f}.
\end{eqnarray}
To check the relation between the fractal dimension of FSLE curves and Hurst index $H$ and diffusivity coefficient $\kappa$, we simply measured the average length of curves $\langle S \rangle$ bounded by a semi-circle with radius $R$. For fractal curves the relation  $\langle S \rangle \sim R^{d_f}$ is expected. The results for FSLE curves with different driving functions FBM are shown in Fig .(\ref{Fig4}). 

An interesting fact about FSLE curves is that all the results for fractal properties of fractional stochastic Loewner evolution with FWM function as driving function were the same as FBM. This the reason that the results were not depicted in the figures. A second fact is that the fractal dimension $d_f$ for FSLE curves with $H =$const surprisingly is a linear function of the diffusivity parameter $\kappa$ (see inset of the figure (\ref{Fig4})). Note also that for fixed values of $\kappa$, the fractal dimension $d_f(H)$ changes dramatically near the value $H=0.75$. We should note that FSLE curve with $H = 1.0$ is a line with $d_f = 1$.

\section{Conclusion}
In this work we have studied the fractional stochastic Loewner evolution. We considered stochastic Loewner evolution with fractional Brownian motion (fractional Weierstrass-Mandelbrot function) as a driving function. To generate scale invariant curves we used the local fractional integro-differential operator (\ref{New SLE}) instead of the normal differential equation (\ref{SLE}). Based on the theoretical arguments and numerical simulation we showed the FSLE curves have a transition with the Hurst index $H$. For
$H<\frac{1}{2}$ the FSLE curves will not touch the real line and for $H>\frac{1}{2}$ the curve will touch the real line
independent of the value of $\kappa$. The critical value $H=\frac{1}{2}$  corresponds to the ordinary SLE. Next, we have investigated fractal property of the FSLE curves. As a result, we find that the fractal dimension $d_f$, depends on the Hurst index $H$ and the diffusivity exponent $\kappa$. The results in Ref. \cite{hermaan2016} are the best support to our new approach to scaling curves. The Ref. \cite{hermaan2016} suggested that the perimeters of multi-layered and directed percolation clusters at criticality are the scaling limits
of the Loewner evolution of an anomalous Brownian motion with exponent $0<\alpha<1$. Investigating statistical models with a similar properties could be interesting.
\section{Acknowledgement}
The authors would like to thank Mohammad Ali Rajabpour for the helpful discussions and suggestions. This work	would	not	have 	been 	possible	without early discussion with MAR. We also appreciate providing
computing facilities of Shiraz University.

\end{document}